\documentclass[
aps,prd,onecolumn,amsmath,amssymb]{revtex4}

\usepackage{graphicx}
\usepackage{dcolumn}
\usepackage{bm}

\usepackage{color}
\usepackage{ulem}

\newcommand{\tcomm}[1]{\textcolor{green}{#1} }

\begin{document}

\preprint{APS/123-QED}

\title{Multi-scale Renormalization Group Methods for Effective Potentials with Multiple Scalar Fields}
\author{T.G.~Steele}
\author{Zhi-Wei Wang}%
\affiliation{%
Department of Physics and
Engineering Physics, University of Saskatchewan, Saskatoon, SK,
S7N 5E2, Canada
}%
\author{D.G.C.~McKeon}
\affiliation{%
Department of Applied Mathematics, University of Western Ontario, London, ON, N6A 5B7, Canada
}%
\affiliation{%
Department of Mathematics and Computer Science, Algoma University, Sault Ste.~Marie, ON,
P6A 2G4, Canada
}%

\begin{abstract}
Multi-scale renormalization group (RG) methods are reviewed and applied to the analysis of the effective potential for radiative symmetry breaking with multiple scalar fields, allowing an extension of the Gildener \& Weinberg (GW) method beyond the weak coupling limit. 
A model containing two interacting real scalar fields is used to illustrate multi-scale RG methods and the multi-scale RG functions of this model are calculated to one-loop order for the $\beta$ function and two-loop order for the anomalous mass dimension. 
The introduction of an extra renormalization scale allows the mapping of the effective potential in this model onto an RG-equivalent form with an $O(2)$ symmetric structure along a particular trajectory in the multiple renormalization-scale space, leading to a simplified form of the effective potential. 
It is demonstrated  that the physical content of the effective potential in the original model, referenced to a single conventional renormalization scale, 
can be extracted from  a particular RG-trajectory that connects to this multi-scale $O(2)$-symmetric form of the effective potential.
Extensions of these multi-scale methods for effective potentials in models containing multiple scalars with $O(M)\times O(N)$ symmetry are also discussed.
\end{abstract}
\maketitle

\section{Introduction}
Following the discovery of the Higgs boson \cite{2012gk,2012gu}, one of the most important outstanding challenges in particle physics is to reveal the underlying mechanism for spontaneous electroweak (EW) symmetry breaking. Amongst the numerous underlying mechanisms, radiative symmetry breaking where spontaneous symmetry breaking can occur through loop (radiative) corrections to the effective potential with a conformal invariant tree-level Lagrangian \cite{Coleman:1973jx} is conceptually appealing since it addresses aspects of the hierarchy and fine-tuning problems \cite{Susskind,Bardeen}. The Higgs mass is protected by classical scale invariance in the radiative Higgs loop corrections \cite{Bardeen,'tHooft:1979bh}, and similar to dimensional transmutation in QCD, leads to natural scale hierarchies in a unification context \cite{Weinberg:1978ym,Hill:2014mqa}. It has been demonstrated by Gildener \& Weinberg (GW) that the above mechanism can be generalized to incorporate arbitrary numbers of elementary scalars beyond the single Higgs EW doublet of the conventional minimal Higgs sector \cite{Gildener}. The GW approach is very useful in models with Higgs portal interactions which lead to natural dark matter predictions \cite{a,b,c,Foot:2010av,Khoze:2013oga,Queiroz:2014yna,Englert:2013gz,Chang:2007ki,Carone:2013wla,AlexanderNunneley:2010nw,Farzinnia:2013pga,Gabrielli:2013hma,
Hambye:2007vf,Heikinheimo:2013cua,Steele:2013fka}. 
However, the GW approach has the limitation that the scalars should be weakly coupled \cite{Gildener,Einhorn,AlexanderNunneley:2010nw}.
This limitation of GW methods precludes analyses  of multi-scalar extensions of interesting radiative EW symmetry-breaking scenarios \cite{Elias:2003zm,Meissner:2006zh}, including those that can describe a 125 GeV Higgs boson \cite{Wang}. For example, extensions of the Coleman-Weinberg (CW)  effective potential \cite{Coleman:1973jx} with an additional heavy Higgs doublet require a large coupling between the two doublets \cite{Hill:2014mqa}.

In this article, we  use multi-scale renormalization group methods \cite{Einhorn,C.Ford} to extend the GW method beyond the weak coupling regime. With the introduction of an extra renormalization scale, we are able to choose a particular trajectory in the multiple renormalization-scale space which results in the GW form of logarithmic terms in the effective potential. The resulting simplification facilitates typical renormalization-group (RG) analyses of effective potentials \cite{Sher:1988mj} and allows higher-loop calculations of the effective potential using only the RG functions of the theory \cite{Elias:2003zm,Elias:2003xp,Hill:2014mqa,Chishtie:2007vd,Wang}. To make connection with the physical content of the theory referenced to a conventional single  renormalization scale, we map the multi-scale couplings onto a physical trajectory in the renormalization-scale space to extract solutions for the physical predictions.  

In Section~\ref{multi_scale_section} we apply the multiple-renormalization scale methods of Ref.~\cite{Einhorn,C.Ford} to a theory of two interacting real scalars, obtaining the multi-scale RG functions and verifying some self-consistency requirements of the approach.    In Section~\ref{effective_potential_section}, we study the  effective potential in the conformal limit of this model to illustrate how the GW method can be extended using multi-scale RG methods.  Generalizations to other models are discussed in Section~\ref{discussion_section}.

\section{Multi-scale Renormalization Group Equation}
\label{multi_scale_section}
Any conformal invariance present in a classical renormalizable field theory is broken by radiative corrections as the process of renormalization inevitably results in the introduction of a non-physical parameter with the dimension of mass. Any change in this parameter must be accompanied by a corresponding change in the quantities that characterize the theory (the couplings, masses, and fields). This results in Green's functions satisfying the RG equation. Satisfying this equation ensures that physical processes are not dependent on the choice of mass scale; one compensates for explicit dependence on the mass scale by having implicit dependence on the mass scale through the couplings, masses and fields that are present \cite{peterman,Gellmann,G.'t Hooft,Weinberg,Callan:1970yg,Symanzik:1970rt,P.M.Stevenson,J.C.Collins1,J.C.Collins2}.

It is apparent that there are widely varying mass scales in nature; the electroweak, strong, grand unified and gravitational mass scales differ by orders of magnitude. This has resulted in a discussion in the literature on how the effective potential (and subsequent spontaneous symmetry breaking it includes) can have different mass scales \cite{Einhorn,C.Ford}. The most convenient approach to the problem is through using the so-called ``minimal subtraction" (MS) approach to renormalization \cite{G.'t Hooft,J.C.Collins1,J.C.Collins2}; this is a mass-independent renormalization scheme that employs dimensional regularization (DR) \cite{M.J.G.Veltman}.

In this MS scheme, a bare coupling $g_B^i$ is dimensionful as it appears in an n-dimensional Lagrangian. If this is a renormalizable scalar coupling in four dimensions, then $g_B^i$ is expanded in powers of the renormalized couplings $g_R^j$ (which are dimensionless)

\begin{equation}
g_B^i=\mu^{-\epsilon}\left(g_R^i+\sum_{\nu=1}^{\infty}\frac{a_v^i\left(g_R^j\right)}{\epsilon^\nu}\right)~~\left(\epsilon=n-4\right)\,.\label{defcoupling}
\end{equation}
The massive parameter $\mu$ is the renormalization scale mentioned above; its contribution to Green's functions cannot be physical and this observation that results in the RG equation.

In what follows, we consider the approach of \cite{Einhorn, C.Ford} to multi-scale problems and replace the single parameter $\mu$ in \eqref{defcoupling} with a series of parameters $\mu_i$  a separate one for each coupling. As none of these parameters are physical, each results in its own RG equation.

For purposes of illustration, we consider a model with two scalars $\phi_1$ and $\phi_2$ in four dimensions possessing the symmetries $\phi_1\leftrightarrow\phi_2$ and $\phi_i\rightarrow -\phi_i$. The simplest renormalizable model with these properties has the action 
\begin{equation}
S=\int d^4x\bigg[\frac{1}{2}\left(\partial_\mu\phi_1\right)^2+\frac{1}{2}\left(\partial_\mu\phi_2\right)^2-\frac{m_B^2}{2}\left(\phi_1^2+\phi_2^2\right)
-\frac{\lambda_B}{4!}\left(\phi_1^4+\phi_2^4\right)-\frac{g_B}{2!2!}\phi_1^2\phi_2^2-\Lambda\bigg]\label{lagrangian}
\end{equation}
where $\Lambda$ is a cosmological term \cite{Kastening:1996nj}.
If one employs DR with different mass scales for the couplings $\lambda_B$ and $g_B$, then

\begin{equation}
\lambda_B=\mu_\lambda^{-\epsilon}\left(\lambda_R+\sum_{\nu=1}^{\infty}\frac{a_\nu}{\epsilon^\nu}\right)\label{lambda}
\end{equation}
\begin{equation}
g_B=\mu_g^{-\epsilon}\left(g_R+\sum_{\nu=1}^{\infty}\frac{b_\nu}{\epsilon^\nu}\right)\label{g_B}
\end{equation}
\begin{equation}
m_B^2=m_R^2\left(1+\sum_{\nu=1}^{\infty}\frac{c_\nu}{\epsilon^\nu}\right)\label{m}
\end{equation}
If $\mu_{\lambda}=\mu_g$, then $a_\nu$, $b_\nu$ and $c_\nu$ are dependent only on $\lambda_R$ and $g_R$ \cite{G.'t Hooft,Weinberg,J.C.Collins1,J.C.Collins2}. However, in general, these functions also acquire a dependence on $l=\log\left[\frac{\mu_g^2}{\mu_\lambda^2}\right]$.

The bare quantities are independent of $\mu_\lambda$ and $\mu_g$. This means that in order for Eqs.~\eqref{lambda}--\eqref{m} to be satisfied, the renormalized quantities depend on $\mu_\lambda$ and $\mu_g$; this dependency results in the expansions 

\begin{align}
\mu_\lambda\frac{\partial\lambda_R}{\partial\mu_{\lambda}}=\sum_{\nu=0}^{\infty}x_{\nu}^{\lambda/\lambda}\epsilon^{\nu}\\
\mu_\lambda\frac{\partial g_R}{\partial\mu_{\lambda}}=\sum_{\nu=0}^{\infty}x_{\nu}^{g/\lambda}\epsilon^{\nu}\\
\mu_\lambda\frac{\partial m_R^2}{\partial\mu_{\lambda}}=\sum_{\nu=0}^{\infty}x_{\nu}^{m/\lambda}\epsilon^{\nu}
\end{align}
with similar expansions for derivatives with respect to $\mu_g$ for these renormalized quantities.

\begin{figure}[ht]
\centering
\includegraphics[scale=0.05]{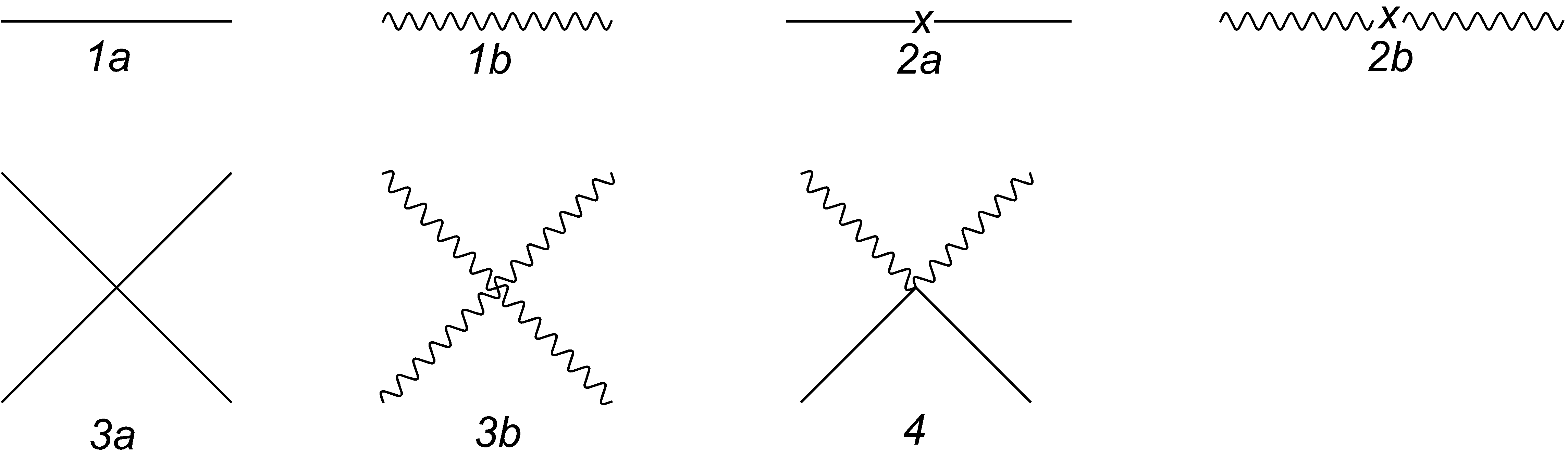}
\caption{Tree level Feynman diagrams are shown for the action \eqref{lagrangian}. The solid line represents $\phi_1$ and wavy line represents $\phi_2$.}
\label{tree}
\end{figure}

\begin{figure}[ht]
\centering
\includegraphics[scale=0.05]{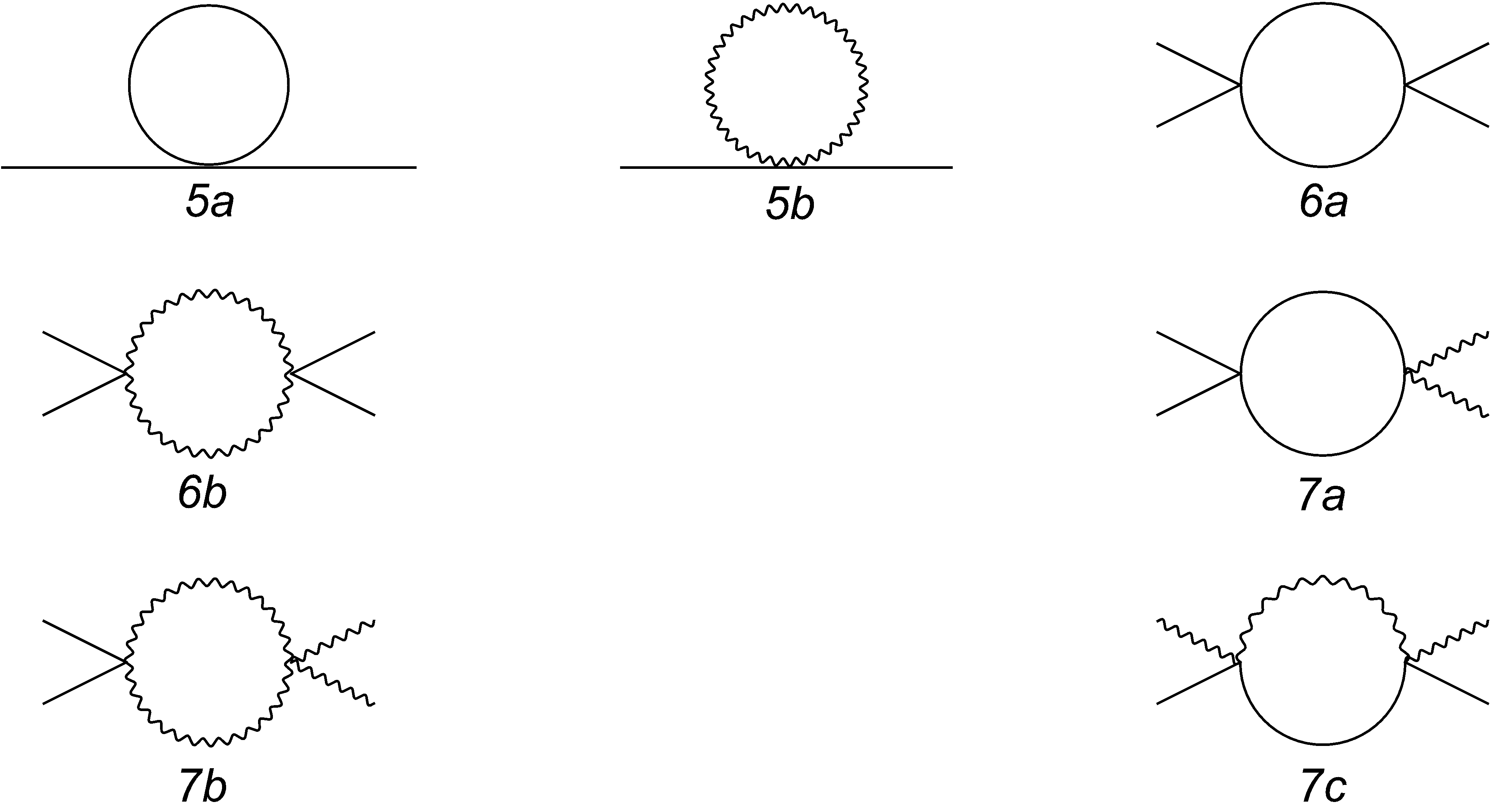}
\caption{One loop level Feynman diagrams for the two point Green functions and four point Green functions.}
\label{loop}
\end{figure}

\begin{figure}[ht]
\centering
\includegraphics[scale=0.05]{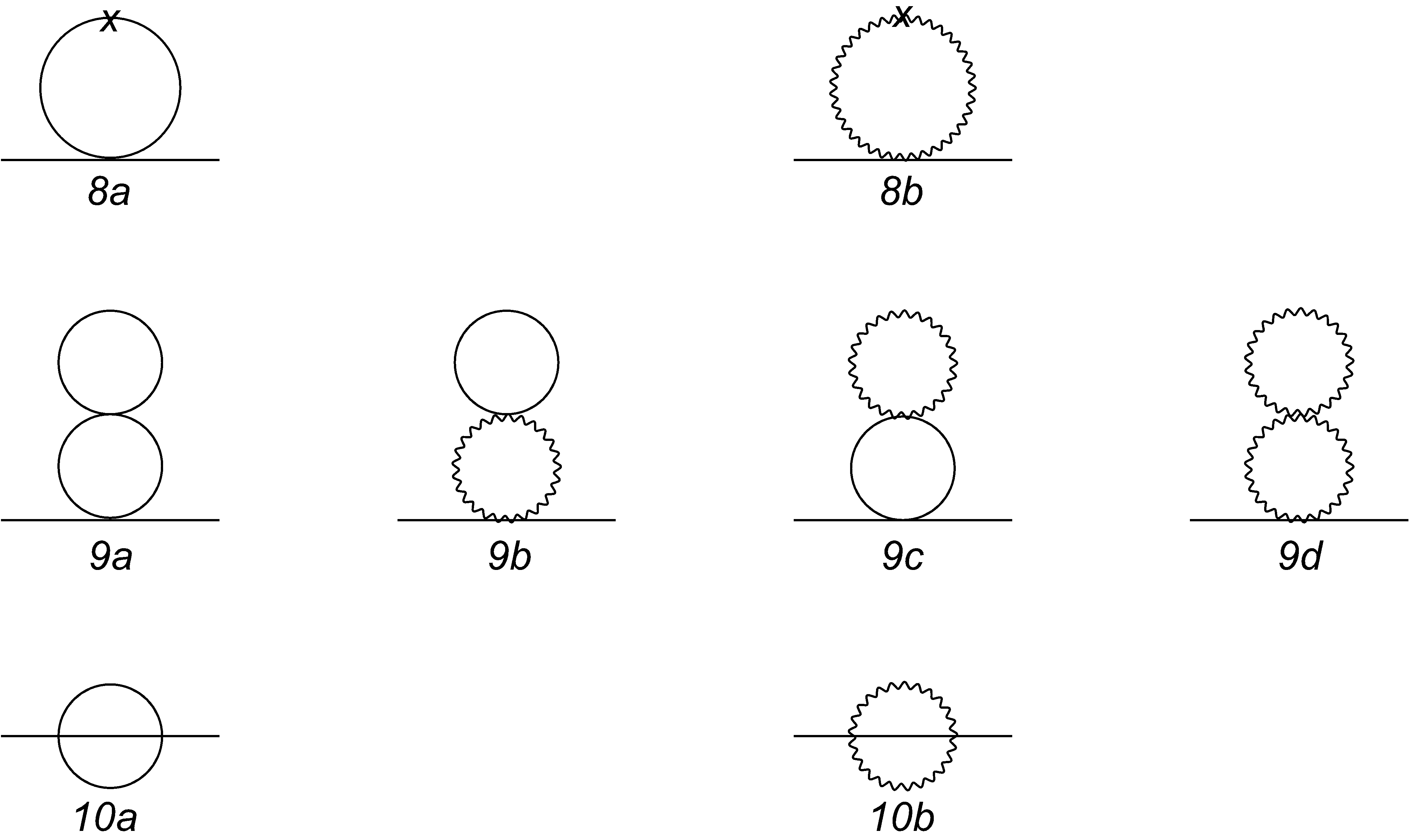}
\caption{Feynman diagrams for two loop order mass anomalous dimension.}
\label{anomalous}
\end{figure}

Following ref.\cite{J.C.Collins1,J.C.Collins2}, we find from Eq.~\eqref{lambda} that
\begin{equation}
\begin{split}
\mu_\lambda\frac{d\lambda_B}{d\mu_\lambda}=0&=\mu_\lambda^{-\epsilon}\Bigg[\left(1+\sum_{\nu=1}^{\infty}\frac{a_{\nu,\lambda}}{\epsilon}\right)\left(\sum_{\nu=0}^{\infty}x_\nu^{\lambda/\lambda}\epsilon^\nu\right)+\left(\sum_{\nu=1}^{\infty}\frac{a_{\nu,g}}{\epsilon^\nu}\right)\left(\sum_{\nu=0}^{\infty}x_\nu^{g/\lambda}\epsilon^\nu\right)\\
&-2\left(\sum_{\nu=1}^{\infty}\frac{a_{\nu,l}}{\epsilon^{\nu}}\right)-\epsilon\left(\lambda_R+\sum_{\nu=1}^{\infty}\frac{a_\nu}{\epsilon^\nu}\right)\Bigg]\label{a}
\end{split}
\end{equation}
If now $x_\nu^{i/j}=0$ for $\nu>1$, then at order $\epsilon$, Eq.~\eqref{a} shows that
\begin{equation}
x_1^{\lambda/\lambda}=\lambda_R\label{b}
\end{equation}
and consequently at order $\epsilon^0$, it follows that
\begin{equation}
x_0^{\lambda/\lambda}=a_1-\lambda_Ra_{1,\lambda}-x_1^{g/\lambda}a_{1,g}\,.\label{c}
\end{equation}
In a similar fashion, we find that
\begin{equation}
\begin{split}
&x_1^{g/g}=g_R\\
&x_0^{g/g}=b_1-g_Rb_{1,g}-x_1^{\lambda/g}b_{1,\lambda}\\
&x_1^{g/\lambda}=x_1^{\lambda/g}=x_1^{m/\lambda}=x_1^{m/g}=0\\
&x_0^{\lambda/g}=-g_Ra_{1,g}\\
&x_0^{g/\lambda}=-\lambda_Rb_{1,\lambda}\\
&x_0^{m/\lambda}=-m_R^2\lambda_R c_{1,\lambda}\\
&x_0^{m/g}=-m_R^2g_R c_{1,g}\label{d}
\end{split}
\end{equation}
If terms of order $\epsilon^{-\nu}\,\left(\nu\geq1\right)$ are considered in Eq.~\eqref{a} and its analogues, we find consistency conditions that are to be satisfied for the functions $a_\nu, b_\nu$ and $c_\nu$. For example, at order $\epsilon^{-1}$, the equation for $\mu_\lambda\frac{d m_B^2}{d\mu_{\lambda}}=0$ results in having
\begin{equation}
\begin{split}
-\lambda_R c_{1,\lambda}c_1+c_{1,\lambda}\left(a_1-\lambda_R a_{1,\lambda}\right)+\lambda_R c_{2,\lambda}&\\
-\lambda_Rb_{1,\lambda} c_{1,g}-2c_{1,l}=0&
\end{split}
\end{equation}
From Eq.~\eqref{b}-\eqref{d} it follows that in the $\epsilon=0$ limit,
\begin{equation}
\begin{split}
\mu_\lambda\frac{\partial\lambda_R}{\partial\mu_\lambda}&=a_1-\lambda_Ra_{1,\lambda}\\
\mu_g\frac{\partial g_R}{\partial\mu_g}&=b_1-g_Rb_{1,g}\\
\mu_\lambda\frac{\partial g_R}{\partial\mu_\lambda}&=-\lambda_Rb_{1,\lambda}\\
\mu_g\frac{\partial\lambda_R}{\partial\mu_g}&=g_Ra_{1,g}\\
\mu_\lambda\frac{\partial m_R^2}{\partial\mu_\lambda}&=-m_R^2\lambda_R c_{1,\lambda}\\
\mu_g\frac{\partial m_R^2}{\partial\mu_g}&=-m_R^2g_R c_{1,g}\label{e}
\end{split}
\end{equation}
If we were to have $\mu_\lambda=\mu_g=\mu$ at the outset, then \cite{G.'t Hooft, J.C.Collins1, J.C.Collins2} we find that
\begin{equation}
\begin{split}
\mu\frac{d\lambda_R}{d\mu}&=a_1-\lambda_Ra_{1,\lambda}-g_Ra_{1,g}\\
\mu\frac{d g_R}{d\mu}&=b_1-g_Rb_{1,g}-\lambda_R b_{1,\lambda}\\
\mu\frac{dm_R^2}{d\mu}&=-m_R^2\left(\lambda_Rc_{1,\lambda}+g_R c_{1,g}\right) \,.\label{f}
\end{split}
\end{equation}
We now will follow ref.\cite{J.C.Collins1} and compute the coefficients $a_1, b_1, c_1$ to the second order in the couplings.  The relevant Feynman diagrams are shown in Figs.~\ref{tree}--\ref{anomalous}.  The results keeping terms of order $\frac{1}{\epsilon}, \frac{1}{\epsilon^2}$ are shown in Table \ref{results} and in the  equations below. We express our results in terms of renormalized quantities.

\begin{table}[ht]
\centering
  \begin{tabular}{|| l | l ||}
    	\hline
diag. $\left(1a,b\right)$ & $\frac{i}{k^2-m_R^2}$\\
diag. $\left(2a,b\right)$ & $-i\left(m_B^2-m_R^2\right)$\\
diag. $\left(3a,b\right)$ &  $-i\lambda_B$\\
diag. $\left(4\right)$ & $-ig_B$\\
diag. $\left(6a,b\right)$ & $\frac{-3i\left(\lambda_R^2+g_R^2\right)}{\left(4\pi\right)^2\epsilon}$\\
diag. $\left(7a,b,c\right)$ & $\frac{-2i\left(\lambda_R^2g_R+2g_R^2\right)}{\left(4\pi\right)^2\epsilon}$\\

    \hline
  \end{tabular}
\caption{Feynman diagrams results for tree level and one loop level.}
\label{results}
\end{table}

\begin{equation}
\begin{split}
{\rm diag.} \left(5a,b\right)=&\frac{im_R^2}{\left(4\pi\right)^2\epsilon^2}\left(3\lambda_R^2+2\lambda_Rg_R+7g_R^2\right)+\frac{im_R^2}{\epsilon}\Bigg[\frac{-\left(\lambda_R+g_R\right)}{\left(4\pi\right)^2}\\
&+\frac{1}{\left(4\pi\right)^4}\Bigg(\frac{1}{2}\left(\gamma-1\right)\left(3\lambda_R^2+2\lambda_Rg_R+7g_R^2\right)+\frac{1}{2}\bigg(3\left(\lambda_R^2+g_R^2\right)\log\frac{m_R^2}{4\pi\mu_\lambda^2}\\
&+2\left(\lambda_Rg_R+2g_R^2\right)\log\frac{m_R^2}{4\pi\mu_g^2}\bigg)\Bigg)\Bigg]
\end{split}
\end{equation}
\begin{equation}
{\rm diag. }\left(8a,b\right)=\frac{im_R^2}{\left(4\pi\right)^2}\left[\frac{\left(\lambda_R+g_R\right)^2}{\epsilon^2}+\frac{1}{\epsilon}\left(\frac{\gamma}{2}\left(\lambda_R+g_R\right)^2+\left(\lambda_R+g_R\right)\left(\lambda_R\log\frac{m_R^2}{4\pi\mu_\lambda^2}+g_R\log\frac{m_R^2}{4\pi\mu_g^2}\right)\right)\right]
\end{equation}
\begin{equation}
{\rm diag. } \left(9a,b,c,d\right)=
\frac{-im_R^2}{\left(4\pi\right)^4}\left[\frac{\left(\lambda_R+g_R\right)^2}{\epsilon^2}+\frac{1}{\epsilon}\left(\left(\lambda_R+g_R\right)^2\left(\gamma-\frac{1}{2}\right)+\left(\lambda_R+g_R\right)\left(\lambda_R\log\frac{m_R^2}{4\pi\mu_\lambda^2}+g_R\log\frac{m_R^2}{4\pi\mu_g^2}\right)\right)\right]
\end{equation}
\begin{equation}
\begin{split}
{\rm diag. } &\left(10a,b\right)=
\\
&\frac{i}{\left(4\pi\right)^4}\left[-\frac{m_R^2}{\epsilon^2}\left(\lambda_R^2+3g_R^2\right)+\frac{1}{\epsilon}\left(-\frac{1}{2}p^2\left(\frac{\lambda_R^2}{6}+\frac{g_R^2}{2}\right)+m_R^2\left(-\gamma+\frac{3}{2}\right)\left(\lambda_R^2+3g_R^2\right)-m_R^2\left(\lambda_R^2\log\frac{m_R^2}{4\pi\mu_\lambda^2}+3g_R^2\log\frac{m_R^2}{4\pi\mu_g^2}\right)\right)\right]
\end{split}
\end{equation}

The integrals and renormalization conventions we use are in refs. \cite{J.C.Collins1,J.C.Collins2}.
We found from the one-loop four point function and the one and two-loop two point functions that 
\begin{equation}
\begin{split}
\lambda_B&=\mu_\lambda^{-\epsilon}\left[\lambda_R-\frac{3}{\left(4\pi\right)^2\epsilon}\left(\lambda_R^2+g_R^2\right)\right]\\
g_B&=\mu_g^{-\epsilon}\left[g_R-\frac{2}{\left(4\pi\right)^2\epsilon}\left(\lambda_Rg_R+2g_R^2\right)\right]\\
m_B^2&=m_R^2\left(1-\frac{\lambda_R+g_R}{\left(4\pi\right)^2\epsilon}+\frac{5\left(\lambda_R^2+3g_R^2\right)}{12\left(4\pi\right)^4\epsilon}+\frac{3g_R^2-\lambda_Rg_R}{2\left(4\pi\right)^4\epsilon}\log\left(\frac{\mu_g^2}{\mu_\lambda^2}\right)+\frac{2}{\left(4\pi\right)^4\epsilon^2}\left(\lambda_R^2+\lambda_Rg_R+2g_R^2\right)\right)\label{g}
\end{split}
\end{equation}
and
\begin{equation}
Z=1+\frac{1}{12\left(4\pi\right)^2\epsilon}\left(\lambda_R^2+3g_R^2\right)
\end{equation}
where $Z$ is the wavefunction renormalization.
From Eqs.~\eqref{e} and \eqref{g} we obtain
\begin{equation}
\begin{split}
\beta_\lambda^\lambda&=\mu_\lambda\frac{\partial\lambda_R}{\partial\mu_\lambda}=\frac{3\left(\lambda_R^2-g_R^2\right)}{\left(4\pi\right)^2}\\
\beta_g^\lambda&=\mu_g\frac{\partial\lambda_R}{\partial\mu_g}=\frac{6g_R^2}{\left(4\pi\right)^2}\\
\beta_g^g&=\mu_g\frac{\partial g_R}{\partial\mu_g}=\frac{4g_R^2}{\left(4\pi\right)^2}\\
\beta_\lambda^g&=\mu_\lambda\frac{\partial g_R}{\partial\mu_\lambda}=\frac{2\lambda_R g_R}{\left(4\pi\right)^2}\\
\gamma_\lambda^m&=\mu_\lambda\frac{\partial m_R^2}{\partial\mu_\lambda}=\frac{m_R^2}{\left(4\pi\right)^2}\Bigg[\lambda_R-\frac{5}{6\left(4\pi\right)^2}\lambda_R^2+\frac{\lambda_R g_R}{2\left(4\pi\right)^4}\log\left(\frac{\mu_g^2}{\mu_\lambda^2}\right)\Bigg]\\
\gamma_g^m&=\mu_g\frac{\partial m_R^2}{\partial\mu_g}=\frac{m_R^2}{\left(4\pi\right)^2}\Bigg[g_R-\frac{5}{2\left(4\pi\right)^2}g_R^2-\frac{6g_R^2-\lambda_R g_R}{2\left(4\pi\right)^4}\log\left(\frac{\mu_g^2}{\mu_\lambda^2}\right)\Bigg]\label{RG functions}
\end{split}
\end{equation}
The RG functions should be compared with the standard one-loop RG functions that follow from Eq.~\eqref{f}
\begin{equation}
\begin{split}
\beta^\lambda&=\mu\frac{\partial\lambda_R}{\partial\mu}=\frac{3}{\left(4\pi\right)^2}\left(\lambda_R^2+g_R^2\right)\\
\beta^g&=\mu\frac{\partial g_R}{\partial\mu}=\frac{2}{\left(4\pi\right)^2}\left(\lambda_Rg_R+2g_R^2\right)\\
\gamma^m&=\mu\frac{\partial m_R^2}{\partial\mu}=m_R^2\left(\frac{\lambda_R+g_R}{\left(4\pi\right)^2}-\frac{5}{6}\frac{\lambda_R^2+3g_R^2}{\left(4\pi\right)^2}\right)\label{standard RG functions}
\end{split}
\end{equation}

It is apparent that at one-loop order, the RG functions are independent of $l=\log\left(\frac{\mu_g^2}{\mu_\lambda^2}\right)$; however, in general, the form of the RG functions with couplings $\xi_i=\left(\lambda_R,g_R\right)$ is
\begin{equation}
\beta_j^i=\mu_j\frac{\partial\xi_i}{\partial\mu_j}=\sum_{n=2}^{\infty}\sum_{a=0}^{n-2}\frac{1}{n!}c_{k_1\dots k_n}^{i/j\left(a\right)}l^a\label{A}
\end{equation}
\begin{equation}
\gamma_i^m=\mu_i\frac{\partial m_R^2}{\partial \mu_i}=m_R^2\sum_{n=1}^{\infty}\sum_{a=0}^{n-1}\frac{1}{n!}d_{k_1\dots k_n}^{i\left(a\right)}\xi_{k_1}\dots\xi_{k_n}l^a
\end{equation}
Diagrams with n-loops contribute to $c_{k_1\dots k_n}^{i/j\left(a\right)}$ and $d_{k_1\dots k_n}^{i\left(a\right)}$. These coefficients are related on account of the consistency condition
\begin{equation}
\left[\mu_i\frac{\partial}{\partial\mu_i},\mu_j\frac{\partial}{\partial\mu_j}\right]=0\label{C}
\end{equation}
For example, $d_{k_1,0}^{i\left(0\right)}$ $d_{k_1,k_2}^{i\left(1\right)}$ and $c_{k_1,k_2}^{i/j\left(0\right)}$ are related as can be seen by applying the above equation to $m_R^2$ and keeping only terms bilinear in $\xi_i$:
\begin{equation}
\begin{split}
\left[\mu_\lambda\frac{\partial}{\partial\mu_\lambda},\mu_g\frac{\partial}{\partial\mu_g}\right]m_R^2&=\mu_\lambda\frac{\partial}{\partial\mu_\lambda}\left[m_R^2\left(\frac{g_R}{\left(4\pi\right)^2}-\frac{6g_R^2-\lambda_Rg_R}{\left(4\pi\right)^4}\log\frac{\mu_g}{\mu_\lambda}\right)\right]-\mu_g\frac{\partial}{\partial\mu_g}\left[m_R^2\left(\frac{\lambda_R}{\left(4\pi\right)^2}+\frac{\lambda_Rg_R}{\left(4\pi\right)^4}\log\frac{\mu_g}{\mu_\lambda}\right)\right]\\
&=\frac{m_R^2}{\left(4\pi\right)^4}\left[\left(\lambda_Rg_R+2\lambda_Rg_R+\left(6g_R^2-\lambda_Rg_R\right)\right)-\left(g_R\lambda_R+6g_R^2+\lambda_Rg_R\right)\right]\\
&=0
\end{split}
\end{equation}

In general $c_{k_1\dots k_n}^{i/j\left(a\right)}$ $\left(a > 0\right)$ is fixed by $c_{k_1\dots k_{n-1}}^{i/j\left(b\right)}$ as follows from applying Eq.~\eqref{C} to $\xi_i$; Similarly $d_{k_1\dots k_n}^{i\left(a\right)}$ $\left(a > 0\right)$ is fixed by applying Eq.~\eqref{C} to $m_R^2$.
Furthermore, $c_{k_1\dots k_n}^{i/j\left(0\right)}$ and $d_{k_1\dots k_n}^{i\left(0\right)}$ can be determined from the usual RG functions. To see this, we first note that upon setting $\mu_R=\mu$ Eq.~\eqref{e} and Eq.~ \eqref{f} show that 

\begin{equation}
\begin{split}
\beta^i&=\sum\beta_j^i\\
\gamma^m&=\sum_j\gamma_j^m
\end{split}
\end{equation}
as, in general if
\begin{equation}
\xi_{iR}=\left(\mu_i\right)^{-\epsilon}\left[\xi_i+\sum_{\nu=1}^{\infty}\frac{a_{\nu}^i}{\epsilon^\nu}\right]
\end{equation}
then
\begin{equation}
\beta^i=\left.\mu\frac{\partial\xi_i}{\partial\mu}\right\vert_{\mu=\mu_R}=a_1^i-\sum_j\xi_ja_{1,\xi_j}^i\label{G}
\end{equation}

\begin{equation}
\beta_j^i=\mu_j\frac{\partial\xi_i}{\partial\mu_j}=\delta_j^ia_1^i-\xi_ja_{1,\xi_j}^i\label{H}
\end{equation}
Next, we note that at n-loop order
\begin{equation}
a_1^{i\left(n\right)}=\sum_{k_1\dots k_{n+1}}\frac{1}{\left(n+1\right)!}a_{k_1\dots k_{n+1}}^i\xi_{k_1}\dots\xi_{k_n+1}
\end{equation}
and so by Eq.~\eqref{G}
\begin{equation}
\beta^i=\sum_{n=1}^{\infty}\left(\frac{1}{\left(n+1\right)!}-\frac{1}{n!}\right)\sum_{k_1\dots k_{n+1}}a_{k_1\dots k_{n+1}}^i\xi_1\dots\xi_{k+1}
\end{equation}
This is shows that $a_{k_1\dots k_{n+1}}^{i\left(n\right)}$ can be found when $\mu_i=\mu$ by examining $\beta^i$ and thus by Eqs. \eqref{A} and \eqref{H}, $c_{k_1\dots k_{n+1}}^{i/j\left(0\right)}$ can be found.

Actually integrating Eq.~\eqref{e} to determine $\xi_1$ and $m_R^2$ is complicated beyond one loop order by dependence of the right side on $l$; at one-loop order integration of Eq.~\eqref{RG functions} is feasible.

\section{Multi-Scale Renormalization Group Methods for the Effective Potential}
\label{effective_potential_section}
In this Section we use multi-scale renormalization group methods to simplify analysis of the effective potential. In models with multiple scalar fields the (one-loop) effective potential depends on the eigenvalues of the mass matrix, which can become complicated non-polynomial functions of the fields if there are no simplifying symmetries \cite{Meissner:2006zh,Einhorn}.
 The GW approach addresses this complexity, but  is limited to weakly coupled theories 
\cite{Gildener,Einhorn,AlexanderNunneley:2010nw}
which may preclude studies of interesting symmetry-breaking scenarios and Standard-Model extensions  \cite{Hill:2014mqa,Elias:2003zm,Elias:2003xp,Elias:2003zm,Wang,Steele:2013fka,Meissner:2006zh}.
It is thus desirable to generalize the GW method to avoid the weak-coupling limitation.
The original CW analysis with the Standard Model Higgs scalar field was based on this weak-coupling assumption, leading to a light Higgs mass of approximately $10~{\rm GeV}$ \cite{Coleman:1973jx}.
However, the weak-coupling limit in CW analysis of the Standard Model can be addressed by a full leading-logarithm RG analysis, increasing the Higgs mass prediction \cite{Elias:2003zm}. Similarly, the weak-coupling limitation of the GW  approach, which is the generalization of CW model to incorporate extra scalar fields, can be addressed by multi-scale RG methods.

For the model introduced in Section \ref{multi_scale_section}, the multi-scale RG method requires two renormalization scales $\mu_\lambda$ and $\mu_g$ (i.e., one for each coupling) \cite{Einhorn}. In the limit when $\mu_\lambda=\mu_g$ the multi-scale method reverts to conventional single-scale RG techniques.

The presence of a second renormalization scale gives the freedom to choose a special relationship between $\mu_\lambda$ and $\mu_g$ to constrain the couplings to satisfy $\lambda=3g$,  leading to an $O(2)$-symmetric version of the Lagrangian \eqref{lagrangian} and resulting in a simplified logarithmic term in the one-loop level effective potential. By contrast to  conventional (single-scale) RG methods where  the $\lambda=3g$ constraint may be satisfied at a certain fixed scale, with multiple scales the constraint will be satisfied along a particular trajectory in the $\mu_\lambda, \mu_g$ plane.  Because the effective potential will satisfy a modified RG equation along this $O(2)$-symmetric trajectory, we can exploit RG methods to analyze the effective potential along this trajectory.   We can then use RG invariance to map the predictions from the symmetric trajectory to obtain the physical content of the model for a conventional (single) renormalization scale.   Thus instead of a complicated and possibly intractable effective potential in the single-scale RG approach, we simplify the analysis of the effective potential on the symmetric RG trajectory and the challenges are shifted to obtaining the multi-scale RG functions on the symmetric RG trajectory and the process of mapping back to the physical single-scale trajectory. However, because this mapping is governed by both the single- and multi-scale RG functions, the procedure is well-established and has no inherent challenges. It should be noted that the multiple renormalization scales can also be attached to the kinetic terms \cite{Ford:1991hw}. However, the implementation of this approach in a generalized $O\left(M\right)\times O\left(N\right)$ model discussed later will be difficult since it does not provide enough renormalization scales, which leads to insufficient freedom to choose a special relationship between the scales to constrain the couplings. Thus, the approach in Ref.~\cite{Einhorn} will be employed, with each renormalization scale attached to one coupling.
 

We begin by studying the conformal (massless) version of the model given in Eq.~\eqref{lagrangian} with multiple renormalization 
scales (note that the cosmological term is not needed in the massless theory \cite{Kastening:1996nj}).
Since our primary interest is in radiative symmetry-breaking, we assume that $g>0$ so that there is no conventional symmetry-breaking at tree-level.
The one loop effective potential 
has the form 
\begin{equation}
V_{1}=\frac{1}{2}\int d^nk\log\left[\det\left(k^2I+M\right)\right]
\end{equation}
where we are integrating over Euclidean moment $k$ and the mass matrix $M$ is
\begin{equation}
M=\left(
\begin{array}{cc}
 \frac{1}{2}\left(\lambda_B \phi _1{}^2+g_B\phi _2{}^2\right) & g_B\phi_1\phi _2 \\
 g_B \phi _1\phi _2 & \frac{1}{2}\left(\lambda_B \phi _2{}^2+g_B\phi _1{}^2\right)
\end{array}
\right)\,.
\end{equation}
The results of the loop integration can be expressed in terms of the eigenvalues $M_+$ and $M_-$  of the matrix $M$
\begin{equation}
V_1=\frac{1}{32\pi^2\epsilon}\left( M_+^2 M_+^{\epsilon/2}+ M_-^2 M_-^{\epsilon/2}\right)
\label{V1_ep}
\end{equation}
where counter-terms that are polynomials in the fields have been omitted.   As emphasized in Ref.~\cite{Einhorn}, we note that the explicit form of the one-loop contributions contains non-polynomial terms contained in $M_\pm$ and thus presents many complications for analysis of the effective potential.  

By contrast with a single renormalization scale, where a variety of ways for introducing the scale will lead to the the same result because there is only a single dimensionless combination possible for the logarithms resulting from \eqref{V1_ep},  a more systematic approach is needed for multiple renormalization.   Including the tree-level contribution $V_0$ into the effective potential
\begin{gather}
V_{\rm eff}=V_0+V_1
\\
V_0=\frac{\lambda_B}{24}\left(\phi_1^4+\phi_2^4\right)+\frac{g_B}{4}\phi_1^2\phi_2^2\,,
\end{gather}
 using the relation \eqref{g} between the bare and renormalized couplings (note that the anomalous field dimension is zero at one-loop order, so the bare and renormalized fields are identical), and expanding to appropriate order in  the couplings and in $\epsilon$ to obtain all finite terms, gives
\begin{gather}
V_{{\rm eff}}=\frac{\lambda_R}{24}\left(\phi_1^4+\phi_2^4\right)+\frac{g_R}{4}\phi_1^2\phi_2^2
+ \frac{1}{64\pi^2}\left(M_+^2\log{ M_+}+M_-^2\log{ M_-}\right)+ A_\lambda\log{\mu_\lambda}+A_g\log{\mu_g}
\label{V_eff_multi}
\\
A_\lambda=-\frac{1}{16\pi^2}\left[  \frac{1}{8} \left(\lambda_R^2-g_R^2\right)\left(\phi_1^4+\phi_2^4\right)+\frac{1}{2}\lambda_R g_R\phi_1^2\phi_2^2\right]
\label{a_lambda}
\\
A_g=-\frac{1}{16\pi^2}\left[  \frac{1}{4} g_R^2\left(\phi_1^4+\phi_2^4\right)+g_R^2\phi_1^2\phi_2^2\right]
\label{a_g}
\end{gather}
where  counter-terms have again been omitted.  
Note that 
\begin{equation}
A_\lambda+A_g=\frac{1}{32\pi^2}\left(M_+^2+M_-^2\right)
\end{equation}
as required to recover the limit of a single renormalization scale.
The quantities $A_\lambda$ and $A_g$ are also completely determined by the multi-scale RG functions \eqref{RG functions}
\begin{gather}
-A_\lambda=\frac{1}{24}\beta_\lambda^\lambda \left(\phi_1^4+\phi_2^4\right)+\frac{1}{4}\beta_\lambda^g\phi_1^2\phi_2^2
\label{a_lambda_rg}
\\
-A_g=\frac{1}{24}\beta_g^\lambda \left(\phi_1^4+\phi_2^4\right)+\frac{1}{4}\beta_g^g\phi_1^2\phi_2^2
\label{a_g_rg}
\end{gather}
and hence the agreement between Eqs.~\eqref{a_g}, \eqref{a_g_rg} and Eqs.~\eqref{a_lambda}, \eqref{a_lambda_rg} demonstrate that our calculations are self-consistent. 

The typical  strategy for analyzing effective potentials is to choose a renormalization scale where the logarithmic terms are zero (see, e.g., Ref.~\cite{Sher:1988mj}). However, even with multiple renormalization scales this is not possible because of the complicated field dependence of the eigenvalues $M_\pm$, and thus a more sophisticated implementation of the multi-scale RG methods is needed. 
We can simplify the eigenvalues $M_\pm$ by exploiting the additional degree of freedom provided by the multi-scale RG to choose a special relationship between $\mu_\lambda$ and $\mu_g$ to constrain the couplings $\lambda$ and $g$ to fulfill 
\begin{equation}
g\left(\mu_\lambda,\mu_g\right)=\frac{1}{3}\lambda\left(\mu_\lambda,\mu_g\right)\,,
\label{constrain}
\end{equation}
such that the tree-level Lagrangian has  $O(2)$ symmetry.
Then the simplified eigenvalues $M_+'$ and $M_-'$ become
\begin{equation}
M_+'=\frac{1}{2}\lambda\left(\phi _1^2+\phi _2^2\right);M_-'=\frac{1}{6}\lambda\left(\phi _1^2+\phi _2^2\right)\,.
\end{equation}
Using these eigenvalues,   rearranging the logarithms in \eqref{V_eff_multi} via $\log{\mu_g}=\log{\left(\mu_g/\mu_\lambda\right)}+\log{\mu_\lambda}$, and ignoring counter-terms (which could thus contain coefficients with explicit $\log{\left(\mu_g/\mu_\lambda\right)}$ dependence), the one-loop effective potential for the constrained RG trajectory \eqref{constrain} becomes
\begin{equation}
V_{{\rm eff}}=\frac{\lambda_R}{24}\left(\phi_1^2+\phi_2^2\right)^2+ \frac{5\lambda_R^2}{1152\pi^2}\left(\phi_1^2+\phi_2^2\right)^2\log{\left(\frac{\lambda_R\left(\phi_1^2+\phi_2^2\right)}{\mu_\lambda^2}\right)}\,.
\label{V_symmetric}
\end{equation}
Thus along the symmetric RG trajectory, the effective potential assumes the simplified GW form \cite{Gildener}, but with no implicit small-coupling assumptions. 

Finally, by using the scheme transformation \cite{Ford:1991hw}
\begin{equation}
\frac{\lambda _R\left(\mu_\lambda\right)}{\mu_\lambda^2}=\frac{1}{\mu ^2}\,,\label{scheme transformation}
\end{equation}
we obtain the CW form of the effective potential along the symmetric RG trajectory at leading-logatrithm order
\begin{equation}
V_{LL}=V_{tree}+\left(\phi _1^2+\phi _2^2\right)^2\left(BL+CL^2+DL^3+EL^4+\ldots\right)~,~L\equiv \log\left[\frac{\phi _1^2+\phi _2^2}{\mu^2}\right]
\label{V_CW}
\end{equation}
where the coefficients $B,C,D,E\ldots$ are functions of the renormalized coupling $\lambda_R$.
These coefficients can be determined from the renormalization group equation \cite{Elias:2003zm,Elias:2003xp,Hill:2014mqa,Chishtie:2007vd}.
Although Eq.~\eqref{V_CW} superficially resembles an effective potential for an $O(2)$-symmetric $\phi^4$ theory, the imprint of the underlying original theory is contained in the RG functions which must be modified to reflect the effect of both the symmetric RG trajectory \eqref{constrain} and the scheme transformation \eqref{scheme transformation}. 

From Section~\ref{multi_scale_section}, we already know how each coupling runs with the scales $\mu_\lambda,\mu_g$. However, we have chosen a particular  trajectory in the $\mu_\lambda-\mu_g$ plane to  simplify the form of the effective potential and we need to know how the couplings $\lambda,g$ run on this symmetric trajectory. Thus we have: 
\begin{equation}
\begin{split}
\beta^{\lambda}_{\mu }=\frac{d\lambda}{d\log\mu }&=\left(\frac{\partial \lambda}{\partial\log\mu_\lambda}+\frac{\partial \lambda}{\partial \log\mu_g}\frac{d\log \mu_g}{d\log \mu_\lambda}\right)\frac{d\log\mu_\lambda}{d\log\mu }\\
&=\left(\beta _{\lambda}^{\lambda}+\beta ^{\lambda}_{g}\frac{d\log\mu_g}{d\log \mu_\lambda}\right)\frac{d\log\mu_\lambda}{d\log\mu }\label{simplified RG functions1}
\end{split}
\end{equation}
\begin{equation}
\begin{split}
\beta ^{g}_{\mu }=\frac{d g}{d\log\mu }&=\left(\frac{\partial g}{\partial\log\mu_\lambda}+\frac{\partial g}{\partial\log \mu_g}\frac{d\log \mu_g}{d\log \mu_\lambda}\right)\frac{d\log\mu_\lambda}{d\log\mu }\\
&=\left(\beta ^{g}_{\lambda}+\beta _{g}^{g}\frac{d\log\mu_g}{d\log \mu_\lambda}\right)\frac{d\log\mu_\lambda}{d\log\mu }\,.\label{simplified RG functions2}
\end{split}
\end{equation}
For the anomalous dimension we have
\begin{equation}
\begin{split}
\gamma^{\phi_i}_{\mu }=\frac{d\log Z_{\phi _i}}{d\log\mu }&=\left(\frac{\partial \log Z_{\phi _i}}{\partial\log \mu_\lambda}+\frac{\partial \log Z_{\phi _i}}{\partial\log \mu_g}\frac{d\log\mu_g}{d\log \mu_\lambda}\right)\frac{d\log\mu_\lambda}{d\log\mu }\\
&=\left(\gamma _i^{\lambda}+\gamma _i^{g}\frac{d\log\mu_g}{d\log \mu_\lambda}\right)\frac{d\log\mu_\lambda}{d\log\mu }
\end{split}
\end{equation}
where $\frac{d\log\mu_g}{d\log\mu_\lambda}$ is determined from the symmetry constraint \eqref{constrain}
and $\frac{d\log\mu_\lambda}{d\log\mu}$ is determined from Eq.~\eqref{scheme transformation}.
In particular, we find
\begin{equation}
\frac{d \log\mu_g}{d\log \mu_\lambda}=-\frac{\left(\beta _{\lambda}^{\lambda}-3\beta^{g}_{\lambda}\right)}{\left(\beta^{\lambda}_{g}-3\beta _{g}^{g}\right)}\,.
\end{equation}
By using Eq.~\eqref{RG functions}, the above expression can be simplified to
\begin{equation}
\frac{d\log\mu_g}{d\log\mu_\lambda}=-\frac{1}{2}-\frac{\lambda_R}{g_R}+\frac{1}{2}\frac{\lambda_R^2}{g_R^2}\,.
\label{ratio}
\end{equation}
Inputting Eqs.~\eqref{RG functions} and \eqref{ratio} into Eqs.~\eqref{simplified RG functions1} and \eqref{simplified RG functions2}, we have
\begin{equation}
\begin{split}
&\beta^\lambda_\mu=\frac{1}{\left(4\pi\right)^2}\left(6\lambda_R^2-6g_R^2-6\lambda_R g_R\right)\\
&\beta^g_\mu=\frac{1}{\left(4\pi\right)^2}\left(2\lambda_R^2-2g_R^2-2\lambda_R g_R\right)\,.
\end{split}
\end{equation}
We can now impose the constraint \eqref{constrain} that defines the symmetric trajectory to obtain
\begin{gather}
\frac{d\log\mu_g}{d\log\mu_\lambda}=-\frac{1}{2}-\frac{\lambda_R}{g_R}+\frac{1}{2}\frac{\lambda_R^2}{g_R^2}=1
\label{equal_scale}
\\
\beta^\lambda_\mu=\frac{1}{\left(4\pi\right)^2}\frac{10}{3}\lambda_R^2
\\
\beta^g_\mu=\frac{1}{\left(4\pi\right)^2}\frac{10}{9}\lambda_R^2 \,.
\end{gather}
We speculate that the simple relationship \eqref{equal_scale} is a one-loop artifact, and would become non-trivial at higher-loop order.

As outlined in Refs.~\cite{Elias:2003zm,Elias:2003xp,Chishtie:2007vd,Wang}, effective potentials of the $O(2)$ form \eqref{V_CW} are best analyzed by choosing the scale $\mu^2=v^2=\langle\phi_1^2+\phi_2^2\rangle$, which results in predictions of the coupling $\lambda_R(\mu=v)=\lambda_s$ and scalar mass $M_s$ (i.e., the mass matrix is diagonal). It should be noted that vacuum expectation value $v$ is a physical RG-invariant observable.
However, these predictions from the symmetric RG trajectory must be mapped back to physical values in the original theory for a conventional single-renormalization scale.  We describe this process in general, keeping in mind that the one-loop case has a number of simplifications including \eqref{equal_scale} and trivial anomalous dimensions for the fields. Although the effective potential has an $O(2)$-symmetric form along the symmetric trajectory, it is important to recognize that the vacuum configuration of the original theory retains its imprint along the symmetric trajectory through the vacuum angle  $\langle\phi_1\rangle /\langle \phi_2\rangle=\tan\theta$.

First, consider a geometric description of the constraint \eqref{constrain} defining the symmetric RG trajectory.  The multi-scale couplings $g\left(\mu_\lambda,\mu_g\right)$ and $\lambda\left(\mu_\lambda,\mu_g\right)$ can be represented by surfaces parameterized by the renormalization scales $\mu_\lambda$ and $\mu_g$. The constraint \eqref{constrain} represents an intersection of surfaces defining a three-dimensional curve, whose projection onto the $\mu_\lambda,~\mu_g$ plane represents the symmetric RG trajectory.  The single-scale limit $\mu_\lambda=\mu_g$ defines the physical RG trajectory, and  the symmetric and physical trajectories intersect at the scale $\mu_\lambda=\mu_g=\mu^*$ (questions related to the existence of $\mu^*$ are discussed below).

Consider the physical content of the original theory parameterized by the RG-invariant vacuum configuration of the fields ($\phi_1=v_1$, $\phi_2=v_2$, $v^2=v_1^2+v_2^2$),  scalar mass spectrum ($M_{\phi_1}$, $M_{\phi_2}$), and couplings referenced to these symmetry-breaking scales
($\lambda_p(v)=\lambda_v$, $g_p(v)=g_v$).  Using the single- and multi-scale RG functions, the physical couplings $\lambda_v$ and $g_v$ are sufficient to determine 
the scale $\mu^*$ by noticing that at the intersection point of the symmetric and physical trajectories, the physical couplings satisfy the constraint $g_p\left(\mu^*\right)=\frac{1}{3}\lambda_p\left(\mu^*\right)$. The symmetric RG trajectories governed by Eqs.~\eqref{simplified RG functions1} and \eqref{simplified RG functions2} are also uniquely determined after we obtain $\mu^*$.
We can thus evolve $\lambda$ along the symmetric trajectory to $\mu=v$ until it reaches the value $\lambda_s$, and the correct value of the physical coupling $\lambda_p$ therefore self-consistently leads to the numerical value $\lambda_s$. 
Thus, the physical couplings of the original theory are now determined. 

A similar procedure is used to obtain the physical mass spectrum.  
The physical-trajectory values of the mass matrix $M^p_{ij}(v)=M^p_{ij}$ (from which the mass eigenvalues  $M_{\phi_1}$, $M_{\phi_2}$ are obtained) must be evolved using the RG equation to the scale $\mu^*$ where the mass matrix becomes diagonal with a single mass scale as required to connect with the $O(2)$-symmetric trajectory. 
 RG evolution of the (diagonal) mass matrix then continues along the symmetric trajectory to $\mu=v$ until it reaches the value $M_s$.  Just as for the couplings, the correct value of the physical-trajectory mass matrix  $M^p_{ij}$ self-consistently leads to $M_s$, and the physical content of the original theory is now completely determined.  

The existence of the scale $\mu^*$ that connects the physical and symmetric trajectories  can be determined by using the single-scale RG functions to evolve the physical couplings from their values at the scale $v$ until the constraint \eqref{constrain} is satisfied.  One might become concerned in the extreme cases $\mu*\ll v$ or $\mu^*\gg v$ that would result from a hierarchy of the couplings $\lambda_v\gg g_v$.  However, in this situation where one of the couplings is small enough to be ignored, the eigenvalues of the mass matrix are simplified, and alternative approaches are needed.   
Finally, we note that the scale $\mu^*$ will always exist in this model, because the single-scale $\beta$ function for the ratio $r=3g/\lambda$
\begin{equation}
\beta_r=\mu\frac{d r}{d\mu}=\lambda\left(-7 r+\frac{4}{3}r^2-r^3\right)~,
\end{equation}
does not change sign, and hence $r$ can be increased monotonically by evolution to a smaller scale or decreased monotonically by evolution to a larger scale.


\section{Discussion}
\label{discussion_section}
The method we described in the above section for the two-scalar model
can be generalized to the case with $O\left(M\right)\times O\left(N\right)$ symmetry. For the generalized scalar sector:
\begin{equation}
L_{s}=\frac{1}{6}\lambda_1\left(\Phi_1^\dagger\Phi_1\right)^2+\frac{1}{6}\lambda_2\left(\Phi_2^\dagger\Phi_2\right)^2+\lambda_3\left(\Phi_1^\dagger\Phi_1\right)\left(\Phi_2^\dagger\Phi_2\right)
\end{equation}
where $\Phi_1^\dagger\Phi_1$ and $\Phi_2^\dagger\Phi_2$ fulfill $O\left(M\right)$ and $O\left(N\right)$ symmetries respectively. The explicit calculation of the effective potential at one loop level in the $\overline{MS}$ scheme is \cite{Gildener, Meissner:2006zh}:
\begin{equation}
\begin{split}
V_{eff}&=\frac{N-1}{256\pi^2}\left(\frac{2}{3}\lambda_1H^2+2\lambda_3\phi^2\right)^2\ln\left(\frac{\frac{2}{3}\lambda_1H^2+2\lambda_3\phi^2}{v^2}\right)\\
&+\frac{M-1}{256\pi^2}\left(\frac{2}{3}\lambda_2\phi^2+2\lambda_3H^2\right)^2\ln\left(\frac{\frac{2}{3}\lambda_2\phi^2+2\lambda_3H^2}{v^2}\right)\\
&+\frac{1}{64\pi^2}F_+^2\ln\left(\frac{F_+}{v^2}\right)+\frac{1}{64\pi^2}F_-^2\ln\left(\frac{F_-}{v^2}\right)
\end{split}
\end{equation}
where $H^2=\Phi_1^\dagger\Phi_1$,  $\phi^2=\Phi_2^\dagger\Phi_2$ and
\begin{equation}
\begin{split}
&F_{\pm}\left(H,\phi\right):=\frac{\lambda_1+\lambda_3}{2}H^2+\frac{\lambda_2+\lambda_3}{2}\phi^2\\
&\pm\sqrt{\left[\frac{\lambda_1-\lambda_3}{2}H^2-\frac{\lambda_2-\lambda_3}{2}\phi^2\right]^2+4\lambda_3^2\phi^2H^2}\,.
\end{split}
\end{equation}
It should be noted that the above expression of the effective potential is calculated by using $Tr\left[M^4\ln\left(\frac{M^2}{v^2}\right)\right]$ where $M^2$ is the $\left(M+N\right)\times\left(M+N\right)$ mass matrix calculated from the Lagrangian and logarithmic terms emerge from the eigenvalues of the mass matrix. By associating each coupling $\lambda_1, \lambda_2, \lambda_3$ with the scale $\mu_{\lambda_1}, \mu_{\lambda_2}, \mu_{\lambda_3}$, we have the freedom to choose a particular trajectory in the renormalization scale space $\mu_{\lambda_1}- \mu_{\lambda_2}-\mu_{\lambda_3}$, as we did in the last section, such that a tree-level $O(N+M)$-symmetric Lagrangian would be obtained:
\begin{equation}
\begin{split}
&\lambda_1\left(\mu_{\lambda_1}, \mu_{\lambda_2}, \mu_{\lambda_3}\right)=\lambda_2\left(\mu_{\lambda_1}, \mu_{\lambda_2}, \mu_{\lambda_3}\right)\\
&\lambda_1\left(\mu_{\lambda_1}, \mu_{\lambda_2}, \mu_{\lambda_3}\right)=3\lambda_3\left(\mu_{\lambda_1}, \mu_{\lambda_2}, \mu_{\lambda_3}\right)\,.
\end{split}
\end{equation}
By using the above relationships, we are able to simplify the four logarithm terms in the above leading order effective potential expression to obtain a GW form \cite{Gildener} of the logarithmic terms along the symmetric RG trajectory. The methodology developed in the previous Section can then be extended to more complicated models with multiple scalars.

In summary, we have applied the multi-scale RG methods originally developed in Refs.~\cite{Einhorn,C.Ford} to the study of effective potentials with multiple scalar fields, using a model with two interacting real scalar fields as a specific example.  The additional degree of freedom represented
by the second renormalization scale allows the identification of a RG trajectory where the effective potential exhibits an $O(2)$ symmetry, aligning with the GW form of the effective potential \cite{Gildener}.  
However, in contrast to the GW method \cite{Gildener,Einhorn,AlexanderNunneley:2010nw}, there  is no explicit requirement for small scalar couplings in this approach. Although the functional form of the effective potential on the symmetric trajectory  has an $O(2)$ symmetry, the imprint of the original theory remains in the vacuum configuration and in the multi-scale RG functions explicitly calculated for our example model in Section~\ref{multi_scale_section}.  
Thus we have exchanged a complicated (and possibly intractable) non-polynomial structure of the effective potential with a single renormalization scale for a simpler $O(2)$-symmetric structure with multiple renormalization scales; at one- or two-loop order, this simplification is offset by only a marginal increase in the complexity of the RG functions resulting from multiple renormalization scales.  
RG evolution of the effective potential along a specific trajectory connecting the multi-scale symmetric and single-scale regimes then allows the 
physical content of the effective potential to be extracted, further emphasizing  the imprint of the original model that determines the RG functions governing evolution along this trajectory. 

\acknowledgements{T.G.S. is 
grateful for financial support from the Natural Sciences and Engineering Research Council of Canada (NSERC).
D.G.C.M. acknowledges a conversation with Roger Macleod.
}

\end{document}